\documentclass[aps,prl,floatfix,twocolumn,showpacs]{revtex4}%
\usepackage{amssymb}
\usepackage{amsmath}
\usepackage{graphicx}
\usepackage{amsfonts}%
\providecommand{\U}[1]{\protect\rule{.1in}{.1in}}
\providecommand{\U}[1]{\protect\rule{.1in}{.1in}}
\providecommand{\U}[1]{\protect\rule{.1in}{.1in}}
\providecommand{\U}[1]{\protect\rule{.1in}{.1in}}
\providecommand{\U}[1]{\protect\rule{.1in}{.1in}}
\providecommand{\U}[1]{\protect\rule{.1in}{.1in}}
\begin{document}
\title{{Mesoscopic admittance of a double quantum dot}}
\author{Audrey Cottet, Christophe Mora and Takis Kontos}
\affiliation{Laboratoire Pierre Aigrain, Ecole Normale Sup\'{e}rieure, CNRS (UMR 8551),
Universit\'{e} P. et M. Curie, Universit\'{e} D. Diderot, 24 rue Lhomond,
75231 Paris Cedex 05, France}
\date{\today}


\begin{abstract}
We calculate the mesoscopic admittance $G(\omega)$ of a double quantum dot
(DQD), which can be measured directly using microwave techniques. This
quantity reveals spectroscopic information on the DQD and is also directly
sensitive to a Pauli spin blockade effect. We then discuss the problem of a
DQD coupled to a high quality photonic resonator. When the photon correlation
functions can be developed along a RPA-like scheme, the response of the
resonator gives an access to $G(\omega)$.

\end{abstract}
\maketitle

The possibility to couple nanoconductors to capacitive gates has been
instrumental for exploring electronic transport in these systems. Applying DC
gate voltages allows one to tune the energies of localized electronic orbitals
to perform the transport spectroscopy of a nanoconductor and reach various
conduction regimes. Gates can also be coupled to AC electric fields, to obtain
e.g. photo-assisted tunneling or charge pumping \cite{ReviewPAT}. Recently,
the mesoscopic admittance $G(\omega)$ of a single quantum dot subject to an AC
gate voltage has been investigated experimentally\cite{GlattliBoys}. The low
frequency limit $G(\omega\rightarrow0)\simeq-i\omega C_{meso}$ can be
interpreted in terms of a mesoscopic capacitance $C_{meso}$ determined by the
circuit geometric capacitances but also by the dot energy spectrum, which sets
the ability of the dot to absorb electrons. This problem has been discussed
theoretically in the regimes of weak \cite{ButtikerCapaMeso, Nigg} and strong
Coulomb interactions \cite{Hamamoto,Splettstoesser}. In a more quantum view,
gates can mediate a coupling between the electrons of a nanocircuit and cavity
photons. This is widely exploited in the context of Circuit-Quantum
ElectroDynamics. Coupling superconducting qubits to a coplanar waveguide
photonic resonator allows an efficient manipulation, coupling and readout of
the qubits \cite{Blais,Wallraff}. In the dispersive regime where a qubit and a
resonator are strongly detuned, the cavity photons experience a frequency
shift which reveals the qubit state. This shift is sometimes discussed in
terms of the qubit mesoscopic capacitance\cite{capaQubit}. The resonant regime
leads to vacuum Rabi oscillations in which the nanocircuit alternatively emits
and reabsorbs a single photon\cite{VacuumRabi}.

Double quantum dots (DQDs) are mesoscopic circuits which can be made out of
e.g. submicronic two dimensional electron gas structures\cite{Petta}, or
top-gated carbon nanotubes\cite{Churchill}. These devices can be used to
elaborate various types of qubits\cite{Burkard,Petta,Hayashi}, and offer
interesting possibilities in the context of Circuit-Quantum
ElectroDynamics\cite{Childress,Cottet'10}. The behavior of a photonic
resonator coupled to a DQD has been recently studied
experimentally\cite{Petersson}. However, on the theoretical side, this problem
has aroused little attention. Besides, the AC gate-biasing of DQDs has been
studied in the context of spin and charge pumping (see e.g. \cite{Riwar} and
Refs. therein) and photo-assisted DC transport(see e.g. \cite{PAT}), but no
theoretical study has been performed in the context of mesoscopic admittance measurements.

In the first part of this paper, we calculate the mesoscopic admittance
$G(\omega)$ of a DQD. We show that this quantity displays a very rich
behavior. In particular, it is directly sensitive to a Pauli spin-blockade
effect\cite{Ono,Churchill}. A measurement of $G(\omega)$ seems an interesting
way to perform the spectroscopy of a DQD, in the context of e.g. a qubit use,
which can forbid invasive DC probes\cite{Cottet'10}. In the second part of
this paper, we discuss the problem of a DQD weakly coupled to a high quality
photonic resonator. The resonator could offer an alternative to direct AC gate
biasing for measuring $G(\omega)$. When the photon correlation functions can
be developed along a RPA-like scheme, both the dispersive and resonant
behaviors of the resonator can be predicted from $G(\omega)$. We briefly
discuss the range of validity of the RPA scheme in the non-interacting limit.
\begin{figure}[ptb]
\includegraphics[width=1.\linewidth]{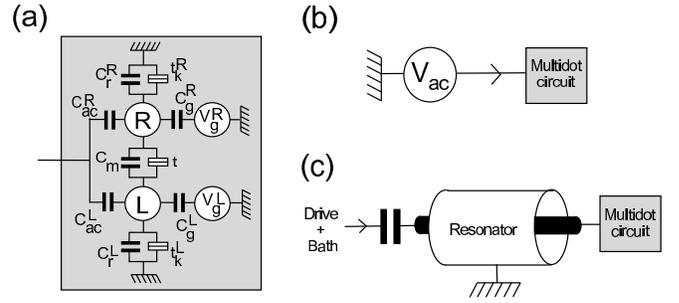}\caption{(a) DQD circuit
considered in this article (b) Configuration used for the measurement of the
DQD mesoscopic admittance (c) Coupling scheme to a photonic resonator}%
\label{Figure1}%
\end{figure}

We first discuss the mesoscopic admittance measurement (Figs.~1a and b). We
consider two single-orbital dots $L$ and $R$ with orbital energies $\xi_{L}$
and $\xi_{R}$, coupled together through a spin-conserving tunnel barrier with
a hoping constant $t$ and a capacitance $C_{m}$. We note $\hat{c}_{d\sigma}$
the annihilation operator associated to an electron with spin $\sigma
\in\{\uparrow,\downarrow\}$ on dot $d\in\{L,R\}$, $\hat{n}_{d\sigma}=\hat
{c}_{d\sigma}^{\dag}\hat{c}_{d\sigma}$, and $\hat{n}_{d}=\hat{n}_{d\uparrow
}+\hat{n}_{d\downarrow}$. Dot $d$ is connected through a tunnel contact to a
grounded reservoir, and connected through a capacitance $C_{g}^{d}$
[$C_{ac}^{d}$] to a DC [AC] bias generator with voltage $V_{g}^{d}$
[$V_{ac}(t)$]. The reservoir states are described by annihilation operators
$\hat{c}_{dk\sigma}$. The full hamiltonian of the circuit writes (up to a term
proportionnal to the identity operator), $\hat{H}_{1}=\hat{H}_{_{DQD}}+\hat
{H}_{l}+\hat{H}_{ac}$ with\cite{VanDerWiel}
\begin{align}
\hat{H}_{DQD} &  =\sum\nolimits_{d,\sigma}(\epsilon_{d}-\sigma\lbrack g\mu
_{B}B/2])\hat{n}_{d\sigma}+\sum\nolimits_{d}\hat{n}_{d}(\hat{n}_{d}%
-1)E_{c}^{d}\nonumber\\
&  +U_{m}\hat{n}_{L}\hat{n}_{R}+t\sum\nolimits_{\sigma}(\hat{c}_{L\sigma
}^{\dag}\hat{c}_{R\sigma}+h.c.)\text{ ,}%
\end{align}%
\[
\hat{H}_{l}=\sum\nolimits_{d,k,\sigma}\left(  [t_{d}\hat{c}_{d\sigma}^{\dag
}\hat{c}_{dk\sigma}+h.c.]+\epsilon_{dk\sigma}\hat{c}_{dk\sigma}^{\dag}\hat
{c}_{dk\sigma}\right)  \text{ ,}%
\]%
\begin{equation}
\hat{H}_{ac}(V_{ac}(t))=\sum\nolimits_{d}e\alpha_{d}\hat{n}_{d}V_{ac}(t)\text{
,}\label{Hac}%
\end{equation}
$\epsilon_{L(R)}=E_{c}^{L(R)}[1-2n_{g}^{L(R)}-2n_{g}^{R(L)}(C_{m}/C_{\Sigma
}^{R(L)})]+\xi_{L(R)}$, $n_{g}^{d}=C_{g}^{d}V_{g}^{d}/e$ and $C_{\Sigma}^{d}$
the total capacitance of dot $d$ \cite{note}. For later use, we define tunnel
rates $\Gamma_{d}=\pi\nu_{0}\left\vert t_{d}\right\vert ^{2}/\hbar$ to the
leads, with $\nu_{0}$ the density of states per spin for reservoir $d$. We
note $\Delta A(t)=\langle\hat{A}-\langle\hat{A}\rangle_{0}\rangle$ with
$\langle\hat{A}\rangle_{0}$ the average value of an operator $\hat{A}$ for
$V_{ac}=0$. From the linear response theory, one finds $\Delta n_{d}%
(\omega)=e(\alpha_{L}\chi_{d,L}(\omega)+\alpha_{R}\chi_{d,R}(\omega
))V_{ac}(\omega)$ with charge correlation functions $\chi_{d,d^{\prime}%
}(t)=-i\theta(t)\langle\lbrack\hat{n}_{d}(t),\hat{n}_{d^{\prime}}]\rangle_{0}%
$. The charge of the capacitor plates connected to $V_{ac}$ writes $\hat
{Q}_{ac}=-\alpha_{L}\hat{n}_{L}e-\alpha_{R}\hat{n}_{R}e+2\lambda_{2}V_{ac}$.
\begin{figure}[ptb]
\includegraphics[width=1.\linewidth]{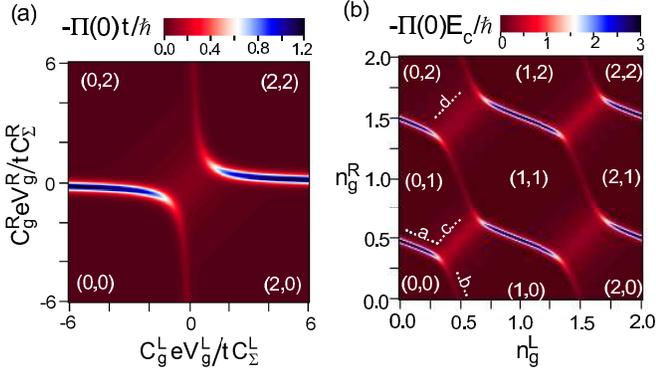}\caption{Response function
$\Pi(0)$ versus the DQD gate voltages in the non-interacting/interacting cases
[panels (a)/(b)]. The white numbers indicate the DQD most stable charge states
\cite{StatesDef}. We have used $\alpha_{L}=-0.1$, $\alpha_{R}=-0.5$ and $B=0$.
In panel (a) we have used $k_{B}T/t=0.1$. In panel (b) we have used
$E_{c}^{L(R)}=E_{c}$, $U_{m}=0.7E_{c}$, $t=0.1E_{c}$, and $k_{B}T=0.02E_{c}$.}%
\label{Figure2}%
\end{figure}Therefore, one obtains
\begin{equation}
\Delta Q_{ac}(\omega)/V_{ac}(\omega)=2\lambda_{2}-(e^{2}/\hbar)\Pi
(\omega)=G(\omega)/(-i\omega)\label{Q}%
\end{equation}
with $\Pi(\omega)=%
{\textstyle\sum\nolimits_{d,d^{\prime}}}
\alpha_{d}\alpha_{d^{\prime}}\chi_{d,d^{\prime}}(\omega)$ and $G(\omega)$ the
admittance of the DQD. The term in $\lambda_{2}$ corresponds to the DQD
response for totally closed quantum dots (i.e. $t_{d}=t=0$). In the low
frequency limit, i.e. $\omega$ much smaller than the characteristic energies
involved in the DQD dynamics (including $\Gamma_{L(R)}$), we obtain
$C_{meso}=2\lambda_{2}-(e^{2}/\hbar)\Pi(0)\in\mathbb{R}$. One can calculate
$\Pi(0)$ from the definition of $\chi_{d,d^{\prime}}(t)$. Alternatively,
assuming that $\hat{n}_{L}$ and $\hat{n}_{R}$ have finite correlation times,
i.e. $\lim_{t\rightarrow+\infty}\chi_{d,d^{\prime}}(t)=0$, one can use%
\begin{equation}
\Pi(0)=\hbar%
{\textstyle\sum\nolimits_{d,d^{\prime}}}
\alpha_{d}\alpha_{d^{\prime}}\partial\left\langle \hat{n}_{d}\right\rangle
_{0}/\partial\epsilon_{d^{\prime}}\label{ZeroFreq}%
\end{equation}
with, assuming that the eigenstates $\left\vert \psi_{i}\right\rangle $ of
$\hat{H}_{DQD}$ (with energies $E_{i}$), are thermally populated\cite{PAT},
\begin{equation}
\left\langle \hat{n}_{d}\right\rangle _{0}=%
{\textstyle\sum\nolimits_{i}}
\left\langle \psi_{i}\right\vert \hat{n}_{d}\left\vert \psi_{i}\right\rangle
\exp(-\beta E_{i})/%
{\textstyle\sum\nolimits_{i}}
\exp(-\beta E_{i})\label{thermal}%
\end{equation}

We first discuss the non-interacting limit, using $E_{c}^{L(R)},U_{m}%
\rightarrow0$, $C_{m}=0$, $B=0$, and $\Gamma_{L(R)}=\Gamma$, which yields
$\epsilon_{d}=-C_{g}^{d}eV_{g}^{d}/C_{\Sigma}^{d}$. In this case, $\Pi
(\omega)$ can be expressed exactly as $\Pi(\omega)=\Pi_{1}(\omega)+\Pi
_{2}(\omega)$, with $\Pi_{1[2]}(\omega)=%
{\textstyle\sum\nolimits_{s\in\{+,-\}}}
\Pi_{s,s[\bar{s}]}(\omega)$,
\begin{equation}
\Pi_{s,s^{\prime}}(\omega)=\frac{4\hbar}{\pi}\int\nolimits_{-\infty}^{+\infty
}d\varepsilon\Gamma f(\varepsilon)g_{s,s^{\prime}}(\omega)/[(\varepsilon
-E_{s})^{2}+\Gamma^{2}] \label{pi12}%
\end{equation}
and $g_{s,s^{\prime}}(\omega)=\lambda_{s,s^{\prime}}^{2}(\varepsilon
-E_{s^{\prime}})/[\left(  \varepsilon-E_{s^{\prime}})^{2}-(\omega
+i\Gamma\right)  ^{2}]$. Here, $\bar{s}$ denotes the sign opposite to $s$. We
use $\lambda_{s,s}=(\alpha_{L}+\alpha_{R}+s(\alpha_{L}-\alpha_{R})\cos
[\theta])/2$, $\lambda_{s,\bar{s}}=-(\alpha_{L}-\alpha_{R})\sin[\theta]/2$,
$\theta=\arctan[2t/(\epsilon_{L}-\epsilon_{R})]$, $E_{\pm}=(\epsilon
_{L}+\epsilon_{R}\pm\Delta_{c})/2$, and $\Delta_{c}=\sqrt{(\epsilon
_{L}-\epsilon_{R})^{2}+4t^{2}}$. In the limit $T=0$ and $\omega=0$, we obtain
$\Pi_{1}(0)=-2\hbar%
{\textstyle\sum\nolimits_{s}}
\lambda_{s,s}^{2}\nu_{s}$ with $\nu_{s}=\Gamma/\pi\lbrack E_{s}{}^{2}%
+\Gamma^{2}]$ the DQD partial density of states (DOS) corresponding to state
$s$ dressed by the leads. This result is reminiscent from the non-interacting
single quantum dot case \cite{ButtikerCapaMeso,remark} where the dot DOS plays
a crucial role. The term $\Pi_{2}(\omega)$ is more specific to the DQD\ case
and is not simply related to $\nu_{s}$. It is finite when $\alpha_{L}%
\neq\alpha_{R}$, i.e. when $V_{ac}$ induces different renormalizations of the
levels $\epsilon_{L}$ and $\epsilon_{R}$. Processes which involve electronic
transfers between the two dots thus contribute crucially to $\Pi_{2}(\omega)$.
We now focus on the limit $0<\Gamma\ll k_{B}T\ll t$. At low frequencies, we
obtain from Eq.(\ref{pi12})%
\begin{equation}
\Pi_{1}(0)=-\beta\hbar\sum\nolimits_{s}\lambda_{s,s}^{2}\cosh^{-2}[\beta
E_{s}/2]/2 \label{pi1therm}%
\end{equation}%
\begin{equation}
\Pi_{2}(0)=-4\hbar(\alpha_{L}-\alpha_{R})^{2}t^{2}(f(E_{-})-f(E_{+}%
))/\Delta_{c}^{3} \label{pi2therm}%
\end{equation}
with $f(\varepsilon)=1/(1+\exp(\beta\varepsilon))$. These results can also be
obtained from Eqs.(\ref{ZeroFreq}) and (\ref{thermal}). Hence, $\Pi(0)$ does
not depend anymore on $\Gamma$. Figure \ref{Figure2}a shows $\Pi(0)$ versus
the DQD gate voltages. The weak resonant line crossing the $V_{g}^{L}%
=V_{g}^{R}=0$ point is due to $\Pi_{2}(0)$ while the anticrossing lines are
due to $\Pi_{1}(0)$. For $\omega,2t\gg\Gamma$, we obtain $\Pi(\omega)\simeq
\Pi_{2}(\omega)$ with $\Pi_{2}(\omega)\simeq4\hbar(\alpha_{L}-\alpha_{R}%
)^{2}t^{2}(f(E_{-})-f(E_{+}))/\Delta_{c}(\omega^{2}-\Delta_{c}^{2})$. Dot/lead
electron transfers are not relevant anymore because they are too slow, and
$\Pi(\omega)$ shows a resonant behavior due to the internal dynamics of the DQD.

We now discuss the interacting case for $0<\Gamma_{L(R)}\ll k_{B}T\ll t\ll
E_{c}^{L(R)},U_{m}$. The DQD\ stability diagram corresponds to the standard
honeycomb pattern\cite{VanDerWiel}. Figure \ref{Figure2}b shows the variations
of $\Pi(0)$ with $n_{g}^{L(R)}$, calculated from Eqs. (\ref{ZeroFreq}) and
(\ref{thermal}) for $B=0$. \begin{figure}[ptb]
\includegraphics[width=1.0\linewidth]{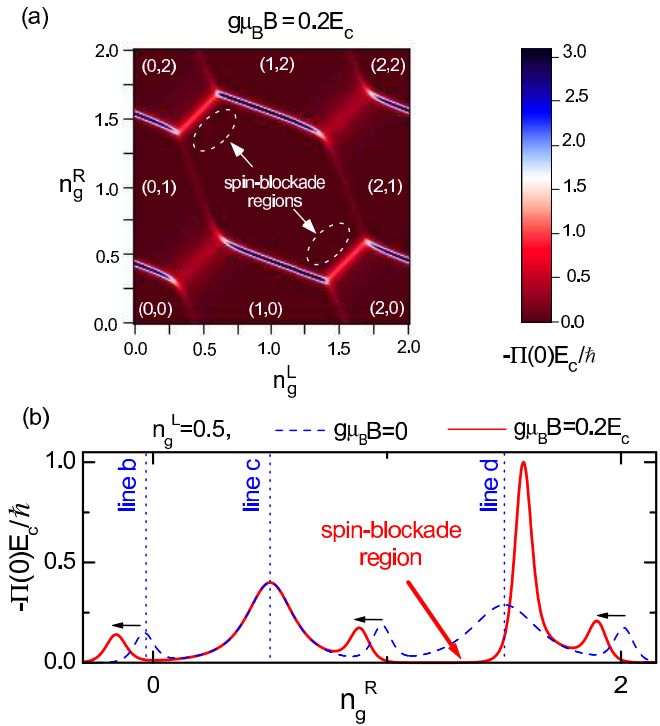}\caption{Effect of a Zeeman
field on $\Pi(0)$. We have used $g\mu_{B}B=0.2E_{c}$ for panel a and the red
full line in panel b, and $g\mu_{B}B=0$ for the blue dashed line in panel b.
The other parameters are the same as in Fig.2.b}%
\label{Figure3}%
\end{figure}Different kinds of resonant lines occur in this graph. The first
kind corresponds to electron transfers between the DQD and a lead, and has a
width set by $T$. For instance, line $a$ corresponds to transitions between
states $(0,0)$ and $(0,\sigma)$, with $\sigma\in\{\uparrow,\downarrow\}$
\cite{StatesDef}. This line can be approximated (away from triple points) as
$\Pi(0)\simeq-\hbar\alpha_{R}^{2}\beta/4\cosh^{2}[\beta\epsilon_{R}/2]$, which
is reminiscent from Eq.(\ref{pi1therm}). Similarly, line $b$ corresponds to
$\Pi(0)\simeq-\hbar\alpha_{L}^{2}\beta/4\cosh^{2}[\beta\epsilon_{L}/2]$. The
second kind of resonances corresponds to electron transfers between the two
dots, in the same $\hat{n}_{R}+\hat{n}_{L}$ subspace. For instance, line $c$
involves resonances between DQD\ states $\left(  \sigma,0\right)  $ and
$\left(  0,\sigma\right)  $, with $\sigma\in\{\uparrow,\downarrow\}$. It can
be approximated by $\Pi(0)\simeq-2\hbar(\alpha_{L}-\alpha_{R})^{2}t^{2}%
/\Delta_{c}^{3}$, which recalls Eq.(\ref{pi2therm}). Line $d$ corresponds to a
resonance between $\left(  0,\uparrow\downarrow\right)  $, $\left(
\uparrow,\downarrow\right)  $ and $\left(  \downarrow,\uparrow\right)  $. It
can be approximated by $\Pi(0)\simeq-4\hbar(\alpha_{L}-\alpha_{R})^{2}%
t^{2}/\Delta_{d}^{3}$ with $\Delta_{d}=\sqrt{(E_{02}-E_{11})^{2}+8t^{2}}$ and
$E_{02}-E_{11}=\epsilon_{R}-\epsilon_{L}+2E_{c}^{R}-U_{m}$. The above
expressions again do not involve the values of the tunnel rates due to
$\Gamma_{L(R)}\ll k_{B}T$. Along line $c$, $\Pi(0)$ reaches a maximum which is
$\sqrt{2}$ higher than along line $d$, because lines $c$ and $d$ involve
resonances between a different number of states. For $\omega\gg\Gamma_{L(R)}$,
using a master equation approach, we find $\Pi(\omega)\simeq-\Pi
(0)\Delta_{c(d)}^{2}/(\omega^{2}-\Delta_{c(d)}^{2})$ along line $c(d)$. The
finite frequency behavior of $\Pi(\omega)$ will be discussed in a more
complete way elsewhere.

We now discuss the effect of a Zeeman field $B$ on $\Pi(0)$ (see Fig.~3). We
use $B>0$ so that $\uparrow$ spins have a lower energy. Lines of type $a$ or
$b$ are shifted by $B$ because they now correspond essentially to a transfer
of $\uparrow$ spins between the dots and leads. However, their height is
almost not modified (except too close to triple points). For a magnetic field
$g\mu_{B}B\sim t$, $\Pi(0)$ cancels in a region where line $d$ was formerly
extending (see Fig.~3b). This is because in this area, the state $\left(
\uparrow,\uparrow\right)  $ becomes the most stable state, and therefore
charge fluctuations between the two dots become impossible. This effect
represents a near-equilibrium version of Pauli spin blockade \cite{Ono}. As a
result, line $d$ is shifted to higher (lower) values of $n_{g}^{R}$
($n_{g}^{L}$), and it reaches a higher maximum which depends strongly on $T$.
Indeed, we obtain for $g\mu_{B}B\ll E_{c}^{L(R)},U_{m}$%
\begin{equation}
\frac{-\Pi(0)}{\hbar(\alpha_{L}-\alpha_{R})^{2}}\simeq\frac{8t^{2}%
+e^{-\beta(\Lambda-g\mu_{B}B)}(\Delta_{d}^{2}\Lambda\beta+2t^{2}(2-\Delta
_{d}\beta)}{\Delta_{d}^{3}(1+\exp[\beta(\Lambda-g\mu_{B}B)])^{2}}
\label{Bfinite}%
\end{equation}
with $\Lambda=(E_{11}-E_{20}+\Delta_{d})/2$. In contrast, line $c$ is not
affected by a magnetic field $g\mu_{B}B\sim t$. Line $c$ is affected by $B$
once $\left(  \uparrow,\uparrow\right)  $ becomes the most stable state near
$n_{g}^{R}=n_{g}^{L}=0.5$, which occurs only for higher values of magnetic
field $g\mu_{B}B\sim U_{m}$ (not shown).

To conclude this first part, mesoscopic admittance measurements appear as an
interesting alternative to charge sensing\cite{Churchill,chargesensing}, for
performing the spectroscopy of quasi-closed multi-quantum-dot systems. We have
mainly discussed the $\Gamma_{L(R)}\ll k_{B}T$\ limit. The frontiers between
the different $(n_{L},n_{R})$ domains can be seen in $\Pi(0)$. In the
interacting case, the parity of the DQD total occupation number can be
determined directly from the difference of amplitude between lines of type $c$
and $d$ obtained at $B=0$, or from the spin blockade effect obtained for
$B\neq0$. The DQD mesoscopic admittance also gives a direct access to
information on the DQD spin state, since spin singlet and triplet states can
be discriminated using spin blockade. At high frequencies $\omega\sim
t\gg\Gamma$, $\Pi(\omega)$ shows resonances due to the internal dynamics of
the DQD. We have disregarded spin and orbital relaxation effects (with rates
denoted $\Gamma_{rel}^{s/o}$), which can be due e.g. to magnetic impurities,
spin-orbit coupling, or phonons. However, assuming $\Gamma_{rel}^{s(o)}%
,\Gamma_{L(R)}\ll k_{B}T$, the results presented here [Eqs.(\ref{pi1therm}) to
(\ref{Bfinite})] will not be affected for $\omega$ much smaller or much larger
than $\Gamma_{rel}^{s(o)},\Gamma_{L(R)}$. For an intermediary value of
$\omega$, the expression of $\Pi(\omega)$ can involve explicitly $\Gamma
_{rel}^{s(o)}$ and $\Gamma_{L(R)}$.

We now consider an experiment where the DQD is\ connected through $C_{ac}^{L}$
and $C_{ac}^{R}$ to an external ($L_{r}$,$C_{r}$) circuit which is a simple
model for a photonic resonator\cite{Blais}. The full circuit hamiltonian
$\hat{H}_{2}$ includes terms in $(C_{r}/2)\hat{V}_{ac}^{2}+(1/2L_{r})\hat
{\Phi}_{ac}^{2}$ and $\lambda_{1}\hat{V}_{ac}+\lambda_{2}\hat{V}_{ac}^{2}$ due
to the resonator and DQD respectively, with $\hat{\Phi}_{ac}$ the flux
operator through the inductance $L_{r}$ and $\hat{V}_{ac}$ the operator
associated to $V_{ac}$. We define the charge operator conjugated to $\hat
{\Phi}_{ac}$ as $\hat{Q}_{ac}=\hat{V}_{ac}/C_{r}^{\prime}$ with $C_{r}%
^{\prime}=C_{r}+2\lambda_{2}$ and the photon annihilation operator $\hat
{a}=-i/\sqrt{2\hbar Z_{r}}\hat{\Phi}_{ac}+\sqrt{Z_{r}/2\hbar}\hat{Q}_{ac}$
with $Z_{r}=\sqrt{L_{r}/C_{r}^{\prime}}$. We assume that the resonator photons
are coupled to an external photonic bath corresponding to the annihilation
operator $\hat{b}$\cite{Clerk}. We finally have
\begin{align}
\hat{H}_{2}  &  =\hat{H}_{DQD}+eV_{rms}\sum\nolimits_{d}\alpha_{d}\hat{n}%
_{d}(\hat{a}+\hat{a}^{\dag})+\lambda_{1}V_{rms}(\hat{a}+\hat{a}^{\dag
})\nonumber\\
&  +\hbar\omega_{r}^{\prime}\hat{a}^{\dag}\hat{a}+\sum\nolimits_{p}\hbar
\omega_{p}\hat{b}_{p}^{\dag}\hat{b}_{p}+\sum\nolimits_{p}(\tau\hat{b}%
_{p}^{\dag}\hat{a}+\tau^{\ast}\hat{a}^{\dag}\hat{b}_{p})\nonumber\\
&  +(\kappa\hat{a}e^{i\omega_{d}t}+\kappa^{\ast}e^{-i\omega_{d}t}\hat{a}%
^{\dag})+\hat{H}_{l} \label{Hq}%
\end{align}
with $\omega_{r}^{\prime}=1/\sqrt{L_{r}C_{r}^{\prime}}$ and $V_{rms}%
=\sqrt{\hbar\omega_{r}^{\prime}/2C_{r}^{\prime}}$. The terms in $\kappa$
account for an external driving of the resonator at frequency $\omega_{d}%
/2\pi$\cite{Blais2}. For simplicity, we study the response of the resonator
through its mean voltage. The linear response theory gives $\Delta
V_{ac}(t)=\operatorname{Re}[G_{\hat{a}+\hat{a}^{\dag},\hat{a}^{\dag}}%
(\omega_{d})\kappa^{\ast}e^{-i\omega_{d}t}]$ with $G_{\hat{A},\hat{B}%
}(t)=-i\theta(t)\langle\lbrack\hat{A}(t),\hat{B}]\rangle_{\kappa=0}$. We can
relate $G_{\hat{a}^{\dag},\hat{a}}$ and $G_{\hat{a}^{\dag},\hat{a}^{\dag}}$ to
$\widetilde{\chi}_{d,d^{\prime}}(t)=-i\theta(t)\left\langle [\hat{n}%
_{d}(t),\hat{n}_{d^{\prime}}]\right\rangle $ by using an equations of motion
approach, which takes into account the stationnarity of $G_{\hat{a}^{\dag
}[\hat{a}],\hat{a}^{\dag}}$. We assume that the self-energy terms $%
{\textstyle\sum\nolimits_{p}}
\left\vert \tau_{p}\right\vert ^{2}/(\hbar\omega\pm\hbar\omega_{p}+i0^{+})$
due to the coupling to the outer photon bath write $-i\hbar\Lambda$, with
$\Lambda>0$, to account simply for the finite quality factor of the resonator.
We obtain the exact relation $G_{\hat{a}^{\dag},\hat{a}}(\omega)=G_{0}%
+G_{0}\omega_{rms}^{2}\tilde{\Pi}(\omega)G_{0}$ with $G_{0}=\left(
\omega-\omega_{r}^{\prime}+i\Lambda\right)  ^{-1}$, $\tilde{\Pi}(\omega)=%
{\textstyle\sum\nolimits_{d,d^{\prime}}}
\alpha_{d}\alpha_{d^{\prime}}\widetilde{\chi}_{d,d^{\prime}}(\omega)$ and
$\omega_{rms}=V_{rms}e/\hbar$. Using an analogous expression for $G_{\hat
{a}^{\dag},\hat{a}^{\dag}}$ and assuming $\Lambda\ll\omega_{r}^{\prime}$, one
finds $G_{\hat{a}+\hat{a}^{\dag},\hat{a}^{\dag}}\simeq G_{\hat{a},\hat
{a}^{\dag}}$. To find the poles of $G_{\hat{a},\hat{a}^{\dag}}$, a
self-consistent approach is necessary\cite{RPAWHY,Skoldberg}. We postulate a
RPA-like approximation $G_{\hat{a},\hat{a}^{\dag}}(\omega)=G_{0}+G_{0}%
\omega_{rms}^{2}\Pi(\omega)G_{\hat{a},\hat{a}^{\dag}}(\omega)$, which yields%
\begin{equation}
G_{\hat{a},\hat{a}^{\dag}}^{-1}(\omega)=G_{0}^{-1}-\omega_{rms}^{2}\Pi(\omega)
\label{7}%
\end{equation}
In the limit where $\hbar\omega_{r}^{\prime}$ and $\hbar\omega_{rms}^{2}%
\Pi(0)$ are both much smaller than the energy scales involved in the DQD
dynamics, Eq.(\ref{7}) gives a dispersive shift of the photonic resonance
frequency, i.e. $\omega_{r}^{tot}\simeq\omega_{r}^{\prime}+\omega_{rms}^{2}%
\Pi(0)$. This result can be recovered by considering a classical parallel
($L_{r}$,$C_{r}$) circuit in parallel with a capacitance $2\lambda_{2}%
-(e^{2}/\hbar)\Pi(0)$ following from Eq. (\ref{Q}). Indeed, assuming
$\omega_{rms}^{2}\Pi(0)\ll\omega_{r}^{\prime}$, we expect free oscillations
with a frequency $(L_{r}[C_{r}^{\prime}-(e^{2}/\hbar)\Pi(0)])^{-1/2}%
\simeq\omega_{r}^{tot}$. For larger values of $\omega_{r}^{\prime}$, in the
general case, the response of the resonator is not simply given by $\Pi
(\omega_{r}^{\prime})$ but by the functional form of $\Pi(\omega)$ [and thus
$G(\omega)$]. For instance, let us use the resonant form $\Pi(\omega
)\simeq\Omega/(\omega^{2}-\Delta^{2})$ obtained previously. One expects an
anticrossing effect when the photonic resonator becomes resonant with the DQD.
From Eq. (\ref{7}), we indeed obtain $\omega_{r,\pm}^{tot}=\left(
\Delta+\omega_{r}^{\prime}\right)  /2\pm\sqrt{A+(\Delta-\omega_{r}^{\prime
})^{2}/4}$ with $A=(\Omega\omega_{rms}^{2})/(\Delta+\omega_{r}^{\prime})$.

In the non-interacting case, the RPA-like approximation of $G_{\hat{a},\hat
{a}^{\dag}}$ can be justified by using a standard diagrammatic perturbation
theory in $\alpha_{L(R)}$. For each order in $\alpha_{L(R)}$, the contribution
to $G_{\hat{a},\hat{a}^{\dag}}$ corresponding to a series of "bubble" diagrams
must be dominant. In principle, an estimation of diagrams at fourth order in
$\alpha_{L(R)}$ already provides a good indication on the validity of the RPA
scheme\cite{Urban}. From a dimensional analysis, the RPA-like development of
$G_{\hat{a},\hat{a}^{\dag}}$ is valid at least in the regime $T=0$ with
$\Lambda,E_{\pm},\hbar\omega_{r}^{\prime},\hbar\omega_{r}^{tot}-\hbar
\omega_{r}^{\prime}\ll\Gamma$. Considering the relevance of the results given
by Eq.(\ref{7}), the RPA scheme is probably valid in a much wider range of
parameters. However, from the fourth order diagrams, it seems crucial to have
$\omega_{r}^{tot}-\omega_{r}^{\prime}$ and $\Lambda$ small, and $\Gamma$
finite, this assertion being difficult to define quantitatively in the general
case\cite{explain}.

As a conclusion for this second part, we have discussed the behavior of a high
finesse photonic resonator coupled to a DQD. When photonic correlation
functions can be developed along a RPA-like scheme, both the dispersive and
resonant behaviors of the resonator reveal information on the DQD admittance.

\textit{We acknowledge fruitful discussions with B. Dou\c{c}ot.}


\begin{thebibliography}{99}                                                                                               %


\bibitem {ReviewPAT}G. Platero and R. Aguado, Phys. Rep. \textbf{395}, 1 (2004).

\bibitem {GlattliBoys}J. Gabelli, et al., Science \textbf{313}, 499 (2006). G.
F\`{e}ve, et al., Science \textbf{316}, 1169 (2007).

\bibitem {ButtikerCapaMeso}M. B\"{u}ttiker, H. Thomas, A. Pr\^{e}tre, Phys.
Lett. A \textbf{180}, 364 (1993); A. Pr\^{e}tre, H. Thomas, M. B\"{u}ttiker,
Phys. Rev. B \textbf{54}, 8130 (1996).

\bibitem {Nigg}S. E. Nigg, R. Lopez, and M. B\"{u}ttiker, Phys. Rev. Lett.
\textbf{97}, 206804 (2006); Z. Ringel, Y. Imry, and O. Entin-Wohlman, Phys.
Rev. B \textbf{78}, 165304 (2008).

\bibitem {Hamamoto}Y. Hamamoto et al. Phys. Rev. B \textbf{81}, 153305 (2010);
C. Mora and K. Le Hur, Nature Physics \textbf{6}, 697 (2010).

\bibitem {Splettstoesser}J. Splettstoesser et al., Phys. Rev. B \textbf{81},
165318 (2010).

\bibitem {Blais}A. Blais et al. Phys. Rev. A \textbf{69}, 062320 (2004).

\bibitem {Wallraff}A. Wallraff et al. Nature \textbf{431}, 162 (2004).

\bibitem {capaQubit}M. A. Sillanp\"{a}\"{a} et al., Phys. Rev. Lett.
\textbf{95}, 206806 (2005), T. Duty et al., Phys. Rev. Lett. \textbf{95},
206807 (2005).

\bibitem {VacuumRabi}A. Wallraff et al., Nature \textbf{431}, 162 (2004).

\bibitem {Burkard}G. Burkard et al., Phys. Rev. B \textbf{59}, 2070 (1999).

\bibitem {Petta}J.R. Petta et al., Science \textbf{309}, 2180 (2005).

\bibitem {Hayashi}T. Hayashi et al. Phys. Rev. Lett. \textbf{91}, 226804 (2003).

\bibitem {Childress}L. Childress, A. S. S\o rensen, and M. D. Lukin, Phys.
Rev. A \textbf{69}, 042302 (2004).

\bibitem {Cottet'10}A. Cottet and T. Kontos, Phys. Rev. Lett. \textbf{105},
160502 (2010).

\bibitem {Petersson}K. D. Petersson et al., Nano Lett., \textbf{10}, 2789 (2010).

\bibitem {Riwar}R.-P. Riwar and J. Splettstoesser, Phys. Rev. B \textbf{82},
205308 (2010).

\bibitem {PAT}R. Ziegler, C. Bruder, and Herbert Schoeller, Phys. Rev. B
\textbf{62}, 1961 (2000).

\bibitem {Ono}K. Ono et al., Science \textbf{297}, 1313 (2002).

\bibitem {Churchill}H. O. H. Churchill et al., Phys. Rev. Lett. \textbf{102},
166802 (2009).

\bibitem {VanDerWiel}W. G. van der Wiel et al., Rev. Mod. Phys. \textbf{75}, 1 (2002).

\bibitem {note}We use $E_{c}^{L(R)}=C_{\Sigma}^{R(L)}e^{2}/2D$, $U_{m}%
=C_{m}e^{2}/D$, $D=C_{\Sigma}^{L}C_{\Sigma}^{R}-C_{m}^{2}$, $\alpha
_{L(R)}=-(C_{ac}^{L(R)}C_{\Sigma}^{R(L)}+C_{ac}^{R(L)}C_{m})/D$, $\lambda_{2}=%
{\textstyle\sum\nolimits_{d}}
(1+\alpha_{d})C_{ac}^{d}/2$, $C_{\Sigma}^{d}=C_{ac}^{d}+C_{g}^{d}+C_{r}%
^{d}+C_{m}$. We use $e>0$ and $A(\omega)=%
{\textstyle\int\nolimits_{-\infty}^{+\infty}}
A(t)\exp(i\omega t)dt$.

\bibitem {remark}In the non-interacting limit, one must assume that geometric
capacitances have a negligible contribution to $G(\omega)\simeq i\omega
(e^{2}/\hbar)\Pi(\omega)$. In principle, one can account for geometric
capacitances by treating $\hat{H}_{DQD}$ at the Hartree
level\cite{ButtikerCapaMeso}. In the Coulomb-blockade limit, our treatment
fully takes into account geometric capacitances.

\bibitem {StatesDef}We note $(s_{L},s_{R})$ a DQD charge state with dot
$d\in\{L,R\}$ in the occupation state $s_{d}\in\{0,1,2\}$, or $\{0,\uparrow
,\downarrow,\uparrow\downarrow\}$ if the spin state is specified.

\bibitem {chargesensing}J. M. Elzerman et al., Nature \textbf{430}, 431
(2004); Y. Hu et al., Nature Nanotech. \textbf{2}, 622 (2007).

\bibitem {Clerk}A. A. Clerk et al., Rev. Mod. Phys. \textbf{82}, 1155 (2010).

\bibitem {Blais2}A. Blais et al., Phys. Rev. A \textbf{75}, 032329 (2007).

\bibitem {RPAWHY}A perturbative treatment at lowest order in $\alpha_{L(R)}$
would give the absurd result $\omega=\omega_{r}^{\prime}$.

\bibitem {Skoldberg}J. Skoldberg et al., Phys. Rev. Lett. \textbf{101}, 087002 (2008).

\bibitem {Urban}D. F. Urban, R. Avriller, A. Levy Yeyati, Phys. Rev. B
\textbf{82}, 121414 (2010).

\bibitem {explain}If the photon and electron linewidths $\Lambda$ and $\Gamma$
both vanish, all fourth order diagrams diverge like $\left(  \omega-\omega
_{r}^{\prime}\right)  ^{-3}$. If $\Gamma$ remains finite while $\Lambda$
vanishes, the double bubble diagram (DBD) keeps a divergence in $\left(
\omega-\omega_{r}^{\prime}\right)  ^{-3}$, while the others diverge like
$\left(  \omega-\omega_{r}^{\prime}\right)  ^{-2}$. Therefore, we expect that
for $\Gamma$ sufficiently large, and $\Lambda$ and $\omega_{r}^{tot}%
-\omega_{r}^{\prime}$ sufficiently small, the DBD will be the dominant fourth
order contribution to $G_{\hat{a},\hat{a}^{\dag}}^{-1}(\omega_{r}^{tot})$.
\end{thebibliography}
\end{document}